\documentclass[aps,pre,twocolumn,showpacs]{revtex4}
\usepackage{amsmath,amsfonts} 
\usepackage{graphicx}
\begin{document}
\title{Co-evolutionary games on networks}
\author{Holger Ebel}
\email{ebel@theo-physik.uni-kiel.de}
\author{Stefan Bornholdt}
\affiliation{Institut f\"ur Theoretische Physik, Universit\"at Kiel,
Leibnizstra\ss{}e 15, D-24098 Kiel, Germany}

\date{August 14, 2002}

\begin{abstract}
We study agents on a network playing an iterated Prisoner's dilemma against their neighbors.
The resulting spatially extended co-evolutionary game exhibits stationary states which are Nash equilibria.
After perturbation of these equilibria, avalanches of mutations reestablish a stationary state.
Scale-free avalanche distributions are observed that are in accordance with calculations from the Nash equilibria and a confined branching process.
The transition from subcritical to critical avalanche dynamics can be traced to a change in the degeneracy of the cooperative macrostate and is observed for many variants of this game.
\end{abstract}
\maketitle

\section{Introduction}

Much research has been devoted to the statistical physics of complex systems with game theoretic interactions recently.
One motivation are economic systems composed of a large number of agents with simple local interactions giving rise to complex global structures and dynamics \cite{arthur:1999}.
In particular, the problem of stability and uniqueness of equilibria in economic systems has been readdressed in the context of the aggregate behavior of individual agents \cite{kirman:1989}.
Game theory \cite{neumann/morgenstern:1953} and the theory of evolutionary games \cite{maynard-smith:1982} provide a sufficient framework for modeling individual interactions whereas spatial structures has to be taken into account to tackle co-evolutionary dynamics of real-world systems \cite{blume-l:1993,nowak/lewenstein:1994,thompson/cunningham:2002}.

Here, we will consider random networks of agents which face a social dilemma or, in physical terms, a frustrated interaction. Imagine a situation where each player can take two actions, say {\em cooperating} or {\em defecting}. The optimal global outcome would be achieved by all players cooperating. But an individual player can gain much more when exploiting the cooperators by defecting. Such a situation is called a social dilemma or a frustrated interaction. The central question is how social order is possible and how cooperative behavior can emerge. Examples for such spatially extended dilemmas are biological networks, where connected plants may or may not decide to share resources  \cite{oborny/bokros:2000}, the analysis of internet congestion  \cite{huberman/lukose:1997}, models for economic communication \cite{miller/rode:2002}, and, of course, many sociological problems from conflict research to public transportation \cite{axelrod:1984,schulz/mueller:1994}.

A simple model system is given by the iterated Prisoner's dilemma (IPD) \cite{axelrod/hamilton:1981,axelrod:1984} with co-evolutionary dynamics.
The Prisoner's dilemma game is probably the most prominent example of a basic model for the emergence of cooperative behavior in social, economic, and biological systems.
It provides a frustrated two-particle interaction and has been extensively studied by physicists, economists, biologists, and mathematicians. 

%%%%%%%
A spatially extended Prisoner's dilemma was first proposed by Axelrod who concluded that territoriality strongly influences the evolution of cooperation \cite{axelrod:1984}.
Extensive
%%%%%%%
work on the spatial Prisoner's dilemma started in 1992 when Nowak and May explored a cellular automaton based on this game on regular lattices.
They and others found complex spatiotemporal dynamics and emergence of cooperation for strategy spaces confined to the strategies {\em defecting} and {\em cooperating} \cite{nowak/may:1992,nowak/may:1993,huberman/glance:1993,nowak/may:1994,nowak/may:1994b,herz:1994}.
For the Prisoner's dilemma on lattices and strategy spaces confined to only {\em cooperating} and {\em defecting} (and Tit-For-Tat in \cite{szabo/droz:2000}), methods from theoretical physics, as Monte-Carlo simulations, percolation theory, the theory of (nonequilibrium) phase transitions, and the concept of self-organized criticality, were used to understand why cooperators or defectors dominate or coexist in the system \cite{szabo/toke:1998,chiappin/oliveira:1999,szabo/droz:2000,vainstein/arenzon:2001,tomochi/kono:2002,lim/jayaprakash:2002}. 
Lindgren and Nordahl introduced players which act erroneously sometimes, allowing a complex evolution of strategies in an unbounded strategy space \cite{lindgren/nordahl:1994}.
Others found that the limitation of a player's memory to the last encounter, which translates to a bounded strategy space, does not provide a significant drawback for the players \cite{hauert/schuster:1997,hauert/schuster:1998,hauert:1999}.
Evolutionary games on networks, again with only two strategies, have been studied to ask questions how spatial organization influences the transition from defecting to cooperating \cite{abramson/kuperman:2001} and how the players themselves may influence the network topology \cite{zimmermann/miguel:2001}.
In the following, we will study the Prisoner's dilemma on a network with a larger but bounded strategy space and co-evolutionary dynamics that lead to Nash equilibria as stationary states. It will be shown both numerically and theoretically that payoff matrix, strategy space, and topology are crucial to answer the question which equilibria will occur and how stable they are. In particular, critical avalanches of mutations are observed for such games and will be explained in detail.

This paper is organized as follows.
In Section \ref{sec_ipd}, the spatially extended iterated Prisoner's dilemma is described as well as its co-evolutionary dynamics. This is followed in Section \ref{sec_ava} by numerical investigations of avalanche dynamics showing three distinct regimes due to changes in payoff matrix and topology. The observed Nash equilibria are described and explained in Section \ref{sec_equ} which enables us to understand the critical value of the control parameter of the payoff matrix. A confined branching process is introduced in Section \ref{sec_bra} clarifying the relaxation mechanism and the emergence of scale-free behavior. Conclusions are drawn in Section \ref{sec_con}.

\section{A co-evolutionary spatially extended IPD}\label{sec_ipd}
We start with a network with $N$ players as nodes where each player plays an iterated Prisoner's dilemma game against each of its neighbors. The Prisoner's dilemma is a two-person game with two possible actions in each encounter. The payoff matrix of the first player for the two strategies {\it cooperating} and {\it defecting} (denoted by $\hat{s}_1$ and $\hat{s}_2$) is given by
\begin{equation}
A=
\begin{pmatrix}
3 & 0 \\
5 & 1 
\end{pmatrix}=(a_{ij})_{i,j \, \in \, \{1,2\}},\label{payoff-matrix}
\end{equation}
with the entries $a_{ij}=\hat{\pi}_1(\hat{s}_i,\hat{s}_j)$. $\hat{\pi}_1(\hat{s}_i,\hat{s}_j)$ is the payoff of player 1 if player 1 plays strategy $\hat{s}_i$ and player 2 plays $\hat{s}_j$. The game is symmetric, i.e.\ $\hat{\pi}_1(\hat{s}_i,\hat{s}_j)=\hat{\pi}_2(\hat{s}_j,\hat{s}_i)$. Therefore, the corresponding payoff matrix of the second player is the transpose of the first player's matrix. The Prisoner's dilemma game in general is defined by the relations
\begin{equation}
a_{12} < a_{22} < a_{11} < a_{21} \text{ and } a_{12}+a_{21} < 2 a_{11}.\label{eq-pdcond}
\end{equation}
Hence, in one encounter of the Prisoner's dilemma, defecting is the strategy that yields the best payoff regardless of the opponent's strategy. This is no longer the case in the iterated game where mutual cooperation is more favorable than both players defecting or switching between defecting and cooperating. That is the reason why this system is a frustrated system. In each encounter, defecting would maximize a player's payoff. But in the long run, when players will anticipate each other's action, cooperation will in general do much better.

\subsection{IPD with memory on a network}
Let us further specify the strategy space and payoff function of the spatial game. A strategy is viewed as a mapping of an agent's ``knowledge'' to an ``action''. ``Knowledge'' of an agent is given by the previous moves the agent can take into account to decide which action it will take next.
%%%%%%%%%
We define the memory length $m$ of a player as the number of these previous moves and
%%%%%%%%% 
confine the memory of the agents to $m=1$, i.e.\ an agent remembers only its opponent player's last action.  If one player encounters another player it has to decide on its first move without any information about the opponent. Accordingly, the opening move is part of the strategy, too, which can be represented as a lookup table or a binary string (Tab.\ \ref{lookup-table}).
 The finite number of moves of one encounter is not known by any agent. In the course of the game, one player has to play against each of its neighbors on the network. Thereafter, its payoff is given by the average payoff per move and neighbor.

\begin{table}\centering
\begin{tabular}{|c|c|}
\hline
History & Action \\
\hline \hline
0 & 1 \\
\hline
1 & 1 \\
\hline
\hline
First move & 1 \\
\hline
\end{tabular}
\caption{\label{lookup-table} Representation of the strategy of one agent with memory $m=1$ (0: defection, 1: cooperation). The agent determined by the above strategy is an unconditioned cooperator. It cooperates no matter whether its opponent has cooperated or defected in the last move.}
\end{table}

The strategy space of a player $i$ consists of up to 8 pure strategies $S_i \subseteq \{0,1,2,3,4,5,6,7\}$ (cf.\ Tab.\ \ref{strat_tab} for definition of the strategies). The pure-strategy space of the game is $S = \underset{i \in I}{\times} S_i$ with the set of the players $I=\{0,1, \cdots ,n\}$. The (pure strategy) payoff function $\pi_i : S \rightarrow \mathbb{R}$ does not depend on the whole pure strategy profile $s=(s_1, \cdots , s_n)$ but only on the strategies of the neighboring nodes $\pi_i=\pi_i(s_i,\text{neigh}(i))$ and, of course, on the payoff matrix of the Prisoner's dilemma game. Here, the set of the neighbors of a node $i$ is denoted $\text{neigh}(i)$. With $\pi(s)=(\pi_1(s), \cdots , \pi_n (s) )$ the above defined game $(S, I, \pi)$ is a finite normal-form game. Such games in general have at least one Nash equilibrium \cite{fudenberg/tirole:1998}. Here, only pure strategies will be considered neglecting possible mixed-strategy equilibria. In this setting, $s_{\text{D}}=(0, \cdots , 0)$ and $s_{\text{TFT}}=(6, \cdots , 6)$ are Nash equilibria for any payoff matrix $A$ obeying (\ref{eq-pdcond}) and for a sufficiently high number of moves, which can be easily verified. The former equilibrium consists of players always defecting, whereas in the latter state each player repeats its opponent's last move (Tit-For-Tat) with a cooperative opening move (cf.\ Tab.\ \ref{strat_tab} ). 

\begin{table}\centering
\begin{tabular}{|c|c|c|c|}
\hline
No.\ & Strategy & Acronym & Bit String \\
\hline \hline
0 & always defect & sD & 000\\
\hline
1 & suspicious anti-Tit-For-Tat & sATFT & 001\\
\hline
2 & suspicious Tit-For-Tat & sTFT & 010 \\
\hline
3 & suspicious cooperate & sC & 011 \\
\hline
4 & generous defect & gD & 100 \\
\hline
5 & generous anti-Tit-For-Tat & gATFT & 101\\
\hline
6 & generous Tit-For-Tat & gTFT & 110 \\
\hline
7 & always cooperate & gC & 111 \\
\hline
\end{tabular}
\caption{\label{strat_tab} The strategy space of each player consists of up to 8 different pure strategies comprehending all possible strategies for a memory of one move. The first lower case letter of the acronym describes the first move: ``s'' for defecting (suspicious) and ``g'' for cooperating (generous). If the strategy is coded as a bit string the assigned numbers correspond to the respective binary numbers.}
\end{table}

\subsection{Co-evolutionary dynamics}
Let us now introduce mutations of a player's strategy. The lookup table determining the strategy is viewed as a bit string of length $2^m+1$, where $m$ is the memory length as defined above. This bit-string will then be mutated during the iteration of the game.

At the beginning, a random network with a given mean degree $\langle k \rangle$ is generated \footnote{The resulting number of links $L=\langle k \rangle N / 2 \gg 1$ (rounded to an integer value) is distributed randomly with equal probability among all $N (N-1) / 2$ possible pairs of nodes leading to a constant probability $\langle k \rangle / (N-1)$ that an arbitrary pair of nodes is connected.}.  The strategies are assigned randomly, too. All agents play against each of their neighbors initially to update their payoffs. Thereafter, the following steps are iterated:
(i) One agent $i$ is chosen randomly and its strategy is mutated from $s_i$ to a strategy $s_i' \in S_i$ picked out at random.
(ii) The mutated agent plays again against its neighbors and its payoff is compared to the former result. The mutation is accepted in the case of a higher payoff, $\pi_i(s_1,s_2,\dots,s_i',\dots,s_N) > \pi_i(s_1,s_2,\dots,s_i,\dots,s_N)$ , and the payoffs of all neighbors are also updated. This corresponds to the assumptions that accepting any mutation is combined with some costs to the player and that mutations occur on a time scale slower than the time scale of the game.
Iteration of this process leads to a stationary state with a fixed strategy distribution.
In the stationary state, no agent can improve its payoff by changing its strategy whereas the other players' strategies remain unchanged. This state corresponds to the game theoretic Nash equilibrium \cite{nash:1950,fudenberg/tirole:1998}.
Note that, for its decisions, no more information than a player's own payoff is required.  

\begin{figure}[t]
\hspace{-1cm}\includegraphics[width=7.5cm]{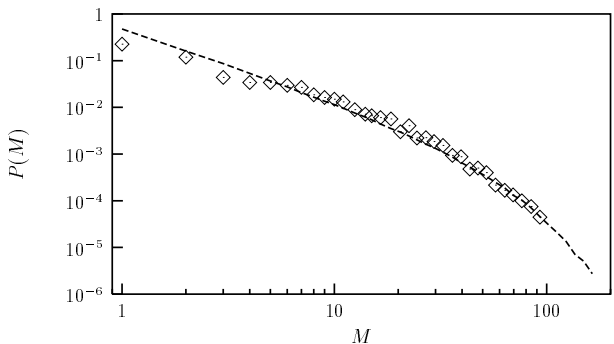} 
\caption{\label{fig_sub}Probability distribution $P(M)$ of avalanche size $M$ for the subcritical case. The avalanche size $M$ is given by the number of mutation events necessary to reestablish an equilibrium.
%%%%%%%%%
With the temptation to defect in the range $3 < a_{21} \leq 4$, only small avalanches are necessary to reestablish the cooperative equilibrium. The open diamonds show data obtained for the spatially extended Prisoner's dilemma averaged over 50 random networks ($N=200$, $\langle k \rangle = 2$, $a_{21}=3.5$). The mechanism of relaxation is a branching process confined to the same topology (dashed curve, $\alpha=0.235$, cf.\ Sec.\ \ref{sec_bra}).}
\end{figure}

\section{Perturbations and Avalanches}\label{sec_ava}
One essential property of evolutionary games is given by the equilibria or the evolutionarily stable states. All stationary states of the game are Nash equilibria. An interesting question is the stability of these equilibria against perturbations. In the following, we will study the dynamics of avalanches of mutation events following a perturbation of the Nash equilibrium. After the system has reached a stationary state, a new strategy is assigned to a random player. The insertion of a suboptimal strategy offers new opportunities for mutations to the perturbed player itself and to its neighbors. Since players are updated randomly, a perturbation leads to an avalanche of mutations until a stationary state is reached again. One quantity of interest is the avalanche size $M$ given by the number of mutations necessary to reestablish the equilibrium and its dependence on the payoff matrix $A$. We will first discuss the numerical results for the case of players on a random network. The second part of this section deals with a Prisoner's dilemma on a ring, which will be the starting point for the theoretical treatment in the next two sections.

\begin{figure}[t]
\hspace{-0.3cm}\includegraphics[width=8.3cm]{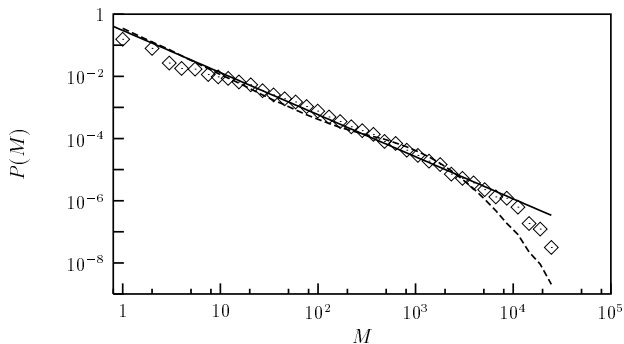} 
\caption{\label{fig_crit}Probability distribution $P(M)$ of avalanche size $M$ (number of mutations events) for the critical case on a random network. The subcritical regime is followed by critical behavior with $4 < a_{21} < 4.70$. The distribution (open diamonds, average over 50 networks with $N=200$, $\langle k \rangle = 2$, and $a_{21}=4.5$) can be well approximated by a power law $P(M) \propto M^{-\gamma}$ with $\gamma=1.39 \pm 0.10$ over three orders of magnitude (solid line). The scale-free behavior can be explained by a confined branching process (dashed curve, $\alpha=0.315$, cf.\ Sec.\ \ref{sec_bra}).}
\end{figure}

\begin{figure}[t]
\hspace{-1cm}\includegraphics[width=8cm]{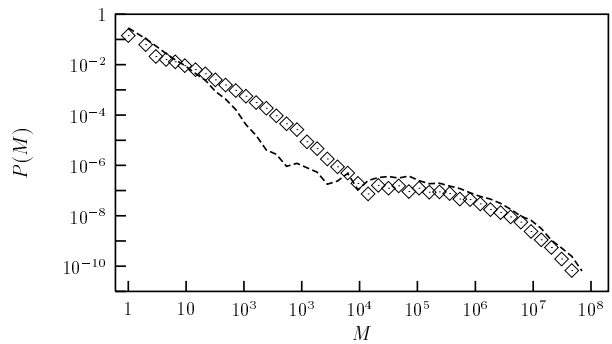} 
\caption{\label{fig_sup}Probability distribution $P(M)$ of avalanche size $M$ (number of mutation events) in the supercritical regime on a random network. For high values of the temptation to defect, $4.7 \leq a_{21} < 6$, a supercritical distribution of the avalanche size is observed (open diamonds, average over 50 networks, $N=200$, $\langle k \rangle = 2$, $a_{21}=4.7$). Again, a confined branching process appears to match the relaxation dynamics well (dashed curve, $\alpha=0.390$, cf.\ Sec.\ \ref{sec_bra}).}
\end{figure}

In the case of sparsely connected random networks, one observes three distinct regimes of the avalanche dynamics, with the temptation to defect $a_{21}$ as control parameter.
For small temptations, $3 < a_{21} \leq 4$, a subcritical regime occurs where large avalanches are suppressed exponentially (Fig.\ \ref{fig_sub}). For $a_{21}>4$, critical behavior occurs with avalanche sizes distributed according to a power law
$
P(M) \propto M^{-\gamma}$
with the scaling exponent $\gamma=1.39 \pm 0.10$ (Fig.\ \ref{fig_crit}) and a cutoff scaling linearly with system size $N$.
This critical regime is followed by a supercritical regime for $4.70 \leq a_{21} < 6$ with an enhanced probability of very large events (Fig.\ \ref{fig_sup}).

Thus, above a critical value of the temptation to defect $a_{21}^{c}=4$, small perturbations of the system lead to long lasting avalanches that affect all players of the whole system with a mean avalanche size that diverges in the thermodynamic limit. The transition from a regime with small avalanches to a critical one with system-wide avalanches is robust in case of moderate changes of the strategy space $S$ and the mean degree $\langle k \rangle$.  It also occurs for smaller strategy spaces $S_i$ with $\text{card}(S_i) \geq 5$ and $\{ 0,6,7 \} \subset S_i$. The qualitative behavior remains even for indefinitely iterated games or with a very different payoff matrix \footnote{
Since $\hat{a}_{12}=\hat{a}_{22}$ the game defined by $\hat{A}$ is not a Prisoner's dilemma in a strict sense (cf.\ Eqn.\ \ref{eq-pdcond}). Furthermore, getting the same payoff in case of a defecting opponent leads to coexistence of cooperative and defective domains in the subcritical regime. 
}, which is sometimes used in the context of the Prisoner's dilemma
\begin{equation}
\hat{A}=\begin{pmatrix}
1 & 0 \\
\hat{a}_{21} & 0
\end{pmatrix}.
\end{equation}
With $S_i=\{0,7\}$ and $\hat{A}$ but quite different evolutionary dynamics, Lim, Chem, and Jayaprakash found critical avalanches on a two-dimensional square lattice, too \cite{lim/jayaprakash:2002}.

The different regimes of relaxation dynamics can be explained by a closer look on the structure of the Nash equilibria involved (Sec.\ \ref{sec_equ}) as well as on the relaxation mechanism which is given by a confined branching process (Sec.\ \ref{sec_bra}). Before we start these considerations in the next two sections, we briefly discuss the case of a co-evolutionary Prisoner's dilemma on a ring, i.e.\ on the regular network where each player's degree is exactly $k_i=2$. Although this is not a very reasonable model for real spatially extended systems, it will give some useful insights and will allow us to calculate some properties of the spatially extended game analytically.

\begin{figure}[t]
\hspace{-1cm}\includegraphics[width=8cm]{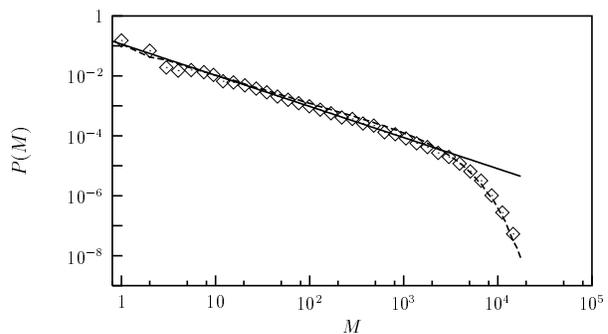} 
\caption{\label{fig_crit_ring}Probability distribution $P(M)$ of avalanche size $M$ (number of mutation events) on a ring in the critical regime. In the range $4.01 \leq a_{21} \leq 4.11$, critical behavior in terms of the avalanche distribution is also observed for the Prisoner's dilemma on a ring ($N=200$, $a_{21}=4.1$). The scale-free distribution $P(M) \propto M^{-\gamma}$ has an exponent of $\gamma=1.04 \pm 0.05$ which is significantly smaller than the scaling exponent observed for the same game on a random network with identical mean degree $\langle k \rangle$. The experimental data agree very well with the behavior of a branching process confined to a ring (dashed curve, $\alpha=0.512$, cf.\ Sec.\ \ref{sec_bra}).}
\end{figure}

Like for random networks, subcritical, critical, and supercritical regimes occur, with subcritical avalanche distributions in the range of $3 < a_{21} \leq 4$ and supercritical behavior for $4.12 \leq a_{21} < 6$. However, this time the critical avalanche distribution has a scaling exponent of $\gamma=1.04 \pm 0.05$ which significantly differs from the exponent obtained for random networks (Fig.\ \ref{fig_crit_ring}).

\section{Nash equilibria and their dependence on the payoff matrix}\label{sec_equ}
The set of possible stationary states of the co-evolutionary Prisoner's dilemma is the set of Nash equilibria which, as has been shown above, contains for all $a_{21} \in (3,6)$ the defective equilibrium $s_{\text{D}}=(0, \dots ,0)$ and the Tit-For-Tat equilibrium $s_{\text{TFT}}=(6, \dots, 6)$. One can also consider the macrostates of the system corresponding to the aggregated behavior of the agents. Identifying cooperative moves with ``spin up'' and defecting moves with ``spin down'' the macroscopic behavior at one instant of time is the magnetization of the system. Thus, the strategy profile $s_{\text{D}}$ of all agents playing strategy $0$ corresponds to the minimal magnetization $-1$ whereas the Tit-For-Tat equilibrium $s_{\text{TFT}}$ leads to the maximal magnetization $+1$.

\subsection{Equilibria on rings and random networks}
Starting with the experimental findings for a ring topology, one observes three regimes in terms of adopted equilibria which are exactly matched by the three different regimes of avalanche dynamics. In the subcritical regime, the stationary states are a mixture of the strategies $6$ and $7$, i.e.\ generous-Tit-For-Tat players and unconditioned cooperators, respectively (Tab.\ \ref{strat_tab}). Only generous Tit-For-Tat prevails in the critical regime. With the onset of supercritical behavior, the defective equilibrium $s_{\text{D}}$ turns up. Its fraction of the equilibria reached by the game grows very fast with further increasing temptation to defect.
The first transition can be explained by a simple calculation of the Nash equilibria. With $a_{21} \leq 4$, the cooperative macrostate is degenerated
%%%%%%%%%%%
in many Nash equilibria
%%%%%%%%%%%
since unconditioned cooperators are stabilized by neighbors with the strategy gTFT. Consider a player $i$ with its neighbors playing $s_{i-1}=7$ (i.e.\ generous cooperate or gC) and $s_{i+1}=6$ (gTFT). Then player $i$ has to find a strategy being a compromise between exploiting the cooperator at $i-1$ and maintaining cooperation with its other neighbor, the smarter Tit-For-Tat player at $i+1$. However, for $a_{21} \leq 4$, there is no such strategy yielding a better payoff than gTFT or even gC. This stabilization of the credulously cooperating agents gives rise to a degeneracy of the cooperative macrostate
%%%%%%%%%
in many different strategy profiles that are Nash equilibria,
%%%%%%%%%%
diverging faster than $2^{2 N / 3}$ with the size of the ring. On the other hand, if $a_{21}>4$ there always exists such a compromise strategy and the degeneracy vanishes. That means that below the critical value $a_{21}^c=4$ the macrostate with magnetization $+1$ is strongly degenerated in many Nash equilibria whereas above $a_{21}^c$ there is only one Nash equilibrium with maximal magnetization left regardless of system size
%%%%%%
($s_{\text{TFT}}$).
The other macrostate with minimal magnetization is never degenerated since $s_{\text{D}}$ is the only Nash equilibrium that leads to such a defective macrostate.
%%%%%%%%%%%%%%%
In case of regular lattices with different numbers of next neighbors $k$, the critical value $a_{21}^c$ is given by
\begin{equation}
a_{21}^c (k) = 4 \frac{2 k +1}{k+3}.
\end{equation}
%%%%%%%%%%%
For example, in the case of a two-dimensional lattice with periodic boundary conditions and a von-Neumann neighborhood, we find subcritical, critical, and supercritical behavior, with $a_{21}^c \approx 5.14$ and a critical exponent of $\gamma= 1.3 \pm 0.2$.
%%%%%%%%%%
Of course, for every payoff matrix $A$ satisfying (\ref{eq-pdcond}) exists a finite value for the number of next neighbors $k$ above which the cooperative macrostate is always degenerated. However, $a_{11}$ and $a_{21}$ can be adjusted to increase this border to arbitrarily large values. Nonetheless, there is a reason why, for every payoff matrix, true critical behavior can only take place in sparsely connected networks which will be discussed in the next section.
Why are only cooperative equilibria observed in both the subcritical and the critical range of $a_{21}$? Looking closer at the way to the equilibrium, there are $(\sigma^2 (\sigma +1) (\sigma-1) ) / 2$ transition probabilities for a player $i$ changing its strategy, with $\sigma := \text{card}(S_i)$. When increasing the temptation to defect $a_{21}$, some of these rules change from $0$ to a finite value and the respective inverse rule vice versa. At the transition to the supercritical regime, where the defective equilibrium is reached for the first time, exactly those rules change which govern the stability of the border between cooperative and defective domains. Below that threshold, the cooperative domains grow and above it defecting strategies can spread.

The situation is slightly different on a random network. Since there are always some nodes with a degree higher than the mean degree $\langle k \rangle$, a small degeneracy of the cooperative macrostate can exist even for $a_{21} > a_{21}^c (\langle k \rangle)$. Moreover, disconnected compounds may be in different equilibria at the same time. The highly connected nodes stabilize the cooperative equilibrium so that even for $a_{21} \lesssim 6$ cooperating strategies predominate. Yet these degeneracy does not compensate for the loss of cooperative equilibrium profiles for temptations larger than $a_{21}^c$ which is the reason for the transition from subcritical to critical behavior. The supercritical phase is again caused by the change of transition rules leading to increasing defective domains with their growth hindered by highly connected cooperative nodes.

\subsection{Nash equilibria and ESS}
As we are dealing with an evolutionary game, the question arises if any of the Nash equilibria is also evolutionarily stable. A strategy profile is called an evolutionarily stable state (ESS) if it is stable against the insertion of a small but finite fraction of mutants playing a different strategy \cite{weibull:1996,hofbauer/sigmund:1998}. Therefore, an ESS is a strict Nash equilibrium or a non-strict Nash equilibrium with the additional condition that other best replies play worse against themselves than the ESS strategy against them.
Note that this concept is formulated for two-person games where two players encounter each other by chance.
In this sense, both Nash equilibria $s_{\text{D}}=(0, \cdots ,0)$ and $s_{\text{TFT}}=(6, \cdots ,6)$ are no ESS for $S_i=\{0, 1,2,3,4,5,6,7 \}$ since other best replies score equally well as the equilibrium strategy (strategies $2$ and $7$, respectively). With respect to strategy spaces reduced by $2$ or $7$, the respective (now strict) Nash equilibrium becomes an ESS. But does this concept of stability apply to spatially extended games? Many approaches to evolutionary stability lead to the equivalence of ESS and strict Nash equilibria. So one may conjecture that in a game with $S_i=\{ 0,1,5,6,7 \}$ the profile $s_{\text{D}}$ should be an ESS since it is a strict Nash equilibrium. Yet, as the experiments show, a small perturbation can cause the system to change from the strict Nash equilibrium $s_{\text{D}}$ to  the non-strict cooperative equilibrium $s_{\text{TFT}}$. Thus, when applying the notion of evolutionary stability to spatially extended systems, one has to keep in mind that things may be different here since it is the local surrounding that decides whether an invader will overthrow the incumbent strategy or will fail instead.

\section{Branching processes as a model of the relaxation process}\label{sec_bra}
Having understood the structure of the Nash equilibria and their connection to the transition between different regimes of avalanche dynamics, the question remains how to explain the distinct form of the probability distributions $P(M)$ and in particular the scaling exponents of the critical regimes. In fact, the relaxation process can be described by a type of branching process which very well predicts the scaling exponents for the different topologies as well as the subcritical and supercritical avalanche distributions.  

\subsection{The Galton-Watson process}
The starting point is a simple branching process, also known as Galton-Watson process, which will be reformulated in terms of mutated agents giving rise to future mutations of other players. Let $Z_n$ be the number of mutated players in the $n$-th generation. Each mutated player can cause other mutations in the next generation, with the probability $p_m$ that its mutation is succeeded by $m$ mutations in the next generation. The stochastic process $(Z_n)_{n \in \mathbb{N}_0}$ is called a branching process of the Galton-Watson type. Note that the number of generations constitutes a time scale completely different from the time scale of the game where at each instant of time one player is chosen to mutate its strategy. With the initial condition $Z_0=1$ and the total progeny $Z:=\sum_{i=0}^{\infty}Z_i$, the quantity of interest is the distribution $P(Z=r)$ of total progeny or, in other words, the avalanche size. So far there is no spatial constraint to this process, i.e.\ $Z_i$ is not bounded by the system size, and mutations independently give birth to new mutations. The probability $p_m$ that a mutation of a player with $k$ neighbors will be followed by $m$ mutations in the next generation is given by
\begin{equation}
p_m = \binom{k+1}{m} \alpha^m (1-\alpha)^{k+1-m},
\end{equation} 
being the simplest choice if a player's mutation can only affect its neighborhood including itself. Using generating functions \cite{harris:1963} we calculate $P(Z=r)$ for this special Galton-Watson process with the same $k_i=k$ for all players to
\begin{equation}
P(Z=r) = \frac{\alpha -1}{\alpha k} \sqrt{\frac{k+1}{2 \pi k}} 
\biggl(\frac{k^{k}}{\alpha (1-\alpha)^k (k+1)^{k+1}} \biggr)^{-r}
r^{-3/2}.\label{exact_pz}
%P(Z=r) = \frac{\alpha -1}{\alpha k} \sqrt{\frac{k+1}{2 \pi k}} 
%\biggl(\frac{(k+1)^{-k}(1-\alpha)^{-k}}{\alpha (k+1) k^{-k}} \biggr)^{-r} r^{-3/2}.\label{exact_pz}
\end{equation}
It is useful to introduce the mean number of a mutation's ``children''
$\bar{m}=\alpha (k+1)=EZ_0$ and approximate (\ref{exact_pz}) for $\bar{m} \lesssim 1$
\begin{equation}
P(Z=r)= C r^{-3/2}  e^{-\frac{r}{r_0}},
\end{equation}
with
\begin{equation}
r_0 = \frac{k+1}{2 k} \quad \frac{1}{(1-\bar{m})^2},
\end{equation}
and a constant
\begin{equation}
C=\frac{\bar{m}-k-1}{k \bar{m}} \sqrt{\frac{k+1}{2 \pi k}}.
\end{equation}
If the expectation of the numbers of descendants approaches  one, i.e $\bar{m} \uparrow 1$, the exponential cutoff diverges with $(1-\bar{m})^{-2}$. The process becomes critical with a scale-free avalanche distribution  $P(Z=r) \propto r^{-3/2}$. If $\bar{m} > 1$, the probability is finite that $Z$ does not converge at all \cite{jagers:1975}. The branching process described above, which has no spatial constraints, is characterized by a subcritical, critical, and supercritical regime of its avalanche dynamics. Although this is very similar to the IPD on a random network, in the case of a ring it yields a wrong scaling exponent of $\gamma=3/2$. Such behavior could be gained equally well from a random walk of the number of mutated sites with drift to a reflecting boundary. In the following, we will show that it is the restriction of the branching process to the network topology that completely explains the dynamics and leads to the correct scaling exponents. 

\subsection{Confined branching processes}

\begin{figure}[t]
\includegraphics[width=8cm]{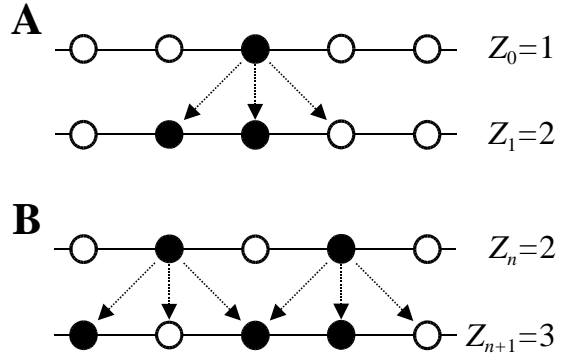} 
\caption{\label{fig_bran}The branching process on a ring. Each node (circle) is occupied by a player with the circle filled if the player's strategy has mutated in the respective generation. 
(A) The initially mutated player causes its neighbors and itself to mutate in the next generation with probability $\alpha$ (arrows). No more players are affected since only its neighborhood and the player itself can experience a different payoff due to the mutation.
(B) The progenies of mutated players in general are not independent of each other. Two mutations can influence the same site making the analytical treatment difficult.  
}
\end{figure}

The confinement of the branching process leads to two effects. First, $Z_n$ will be bounded by the system size $N$; second, the mutation events caused by mutated players are no longer stochastically independent. We will denote a branching process as confined or restricted to a network
(i) if there exists a one-to-one mapping of players and nodes
and (ii) if a mutated player can only give birth to mutations in its neighborhood including itself (Fig.\ \ref{fig_bran} A).
 This corresponds to the fact that if a player changes its strategy only the payoffs of its neighbors and of the player itself will be affected. We assume that each neighbor and the mutated site itself has the same probability $\alpha$ of mutation in the next generation. With the random variable $X_\nu^{(n)}$ being $1$ if the player at node $\nu$ is mutated in generation $n$ and $0$ otherwise, the confined process $(Z_n')_{n \in \mathbb{N}_0}$ is defined by
\begin{equation}   
Z_n' = \sum_{\nu=1}^N X_\nu^{(n)}.
\end{equation}
The probability of a mutation at site $\nu$ in generation $n$ is
\begin{equation}
P(X_\nu^{(n)}=1)= 1 - (1 - \alpha)^{\lambda},\label{eq_px1}
\end{equation}
with
\begin{equation}
\lambda = \sum_{\mu \in \text{neigh}(\nu) \cup \{ \nu \} } X_\mu^{(n-1)}.\label{eq_px2}
\end{equation}
The confined branching process $(Z_n')$ can now be used to calculate the avalanche distributions of the spatially extended Prisoner's dilemma numerically. Applying it to random networks, both the subcritical and supercritical distributions are matched well (dashed curves in Fig.\ \ref{fig_sub} and Fig.\ \ref{fig_sup}). 
The distribution of the confined branching process agrees even better with the experimental data in the critical regime (dashed curves in Fig.\ \ref{fig_crit}). The same is true for the Prisoner's dilemma on a ring (Fig.\ \ref{fig_crit_ring}). In both critical cases, the branching process shows the correct finite-size scaling of the cutoff which is proportional to the system size. Note that the critical regimes of the game have different scaling exponents due to network topology which are both correctly obtained by the confined branching process. The critical exponents depend only on the topology rather than on the parameter $\alpha$ of the process. Therefore, the relaxation mechanism of the spatially extended co-evolutionary Prisoner's dilemma is a confined branching process.   

\begin{table}\centering
\begin{tabular}{|c|c|c|c|}
\hline
Dynamics & $\alpha_{\text{mf1}}$ & $\alpha_{\text{mf2}}$ & $\alpha$\\
\hline \hline
subcritical & $0.290$ & $0.234$ & $0.235$ \\
\hline
critical & $0.340 $ & $0.306 $ & $0.315$\\
\hline
supercritical & $0.320$ & $0.308$ & $0.390$\\
\hline
\end{tabular}
\caption{\label{tab_mf}The branching parameter $\alpha$, determined with mean-field approaches. The parameter $\alpha$, obtained for the experimental distributions of the different regimes on a random network (Figs.\ \ref{fig_sub}, \ref{fig_crit}, \ref{fig_sup}), is compared to mean-field results using a random neighborhood and absorbing stable equilibrium states ($\alpha_{\text{mf1}}$) or averaged over realizations of the game ($\alpha_{\text{mf2}}$).}
\end{table}

Mean-field approaches can be applied successfully to explain the parameter $\alpha$ of the confined branching process in the subcritical and critical regime (Tab.\ \ref{tab_mf}). To calculate a mean-field approximation $\alpha_{\text{mf1}}$ of the branching parameter, the transition probabilities of a mutated agent's neighbors are determined using a random neighborhood for both the player and its neighbors. The structure of the game is taken into account only by assuming that the stable strategies are absorbing states. A second approach is to average the transition probabilities over game realizations numerically, yielding $\alpha_{\text{mf2}}$. Both values, $\alpha_{\text{mf1}}$ and $\alpha_{\text{mf2}}$, agree well with the parameter $\alpha$ obtained from the avalanche distributions of the subcritical and critical regime. This corresponds to the explanation that this transition occurs solely because of the change in the degeneracy of the cooperative macrostate at the critical value $a_{21}^c$. The supercritical case is not matched by the mean-field approaches which may be due to the fact that here the dynamics are governed by local effects, i.e.\ the competitive growth at the boundaries between cooperative and defective domains. The dynamics on a ring topology can be explained by a similar mean-field approach, too, if one assumes that the effective maximal number of a player's descendants is approximately two and not three. This reduction of potential progeny is caused by the strong overlap of the neighborhoods in this regular lattice (Fig.\ \ref{fig_bran} B).

Although the definition of the confined Galton-Watson process is quite intuitive and simple, its analytical treatment is not. The reason is that mutation events has become dependent on each other. Two mutations can affect the same site in the next generation (Fig.\ \ref{fig_bran} B) leading to dependent recursive equations (\ref{eq_px1}, \ref{eq_px2}) for the mutation probability. With the simplification that the $Z_n'$ mutated sites of generation $n$ are randomly distributed over the network, one can shed some light on the critical behavior of the confined branching process. The conditional expectations of the number of mutated players are with this assumption
\begin{equation}
E(Z_{n+1}'|Z_{n}') = \bar{m}  \; Z_{n}' \; ( 1 + \xi )
\end{equation} 
with
\begin{equation}
\xi =  \biggl(\bar{m} \frac{Z_n'}{N}\biggr)^{-1} \biggl[ 1 - \biggl(1- \frac{1}{k+1} \bar{m} \frac{Z_n'}{N}\biggr)^{k+1} - \bar{m} \frac{Z_n'}{N} \biggr].
\end{equation}
If $\xi \ll 1$ and $\bar{m} \approx 1$ the confined process approximately is a martingale for all values of $Z_n'$ and should show critical behavior. For $\bar{m} \ll 1$ the avalanche dynamics are subcritical as the process becomes a supermartingale. With $\bar{m} \gg 1$ obviously resulting in supercritical dynamics, the remaining case of interest is $\bar{m} \approx 1$. In the event of highly connected networks with $\langle k \rangle \gg 1$ the correction $\xi$ is of the order $-1$ suppressing large avalanches. Thus critical avalanche dynamics are expected only for sparsely connected networks, for too strong dependencies of mutation events lead to either subcritical or supercritical distributions of avalanche sizes. 

\section{Conclusions}\label{sec_con}
In this paper, we have 
introduced a spatially extended Prisoner's dilemma game with co-evolutionary dynamics that lead to Nash equilibria as stationary states.
We have shown that critical avalanche dynamics are characteristic for a broad range of these games.
The observed intermittent evolution with sudden avalanches of activity is reminiscent of self-organized criticality \cite{bak:1996,jensen-x:1998}.
Depending on the payoff matrix, subcritical, critical, and supercritical regimes can be observed. Calculating the Nash equilibria and introducing a confined branching process, we were able to quantitatively explain the critical value of the control parameter, i.e.\ the temptation to defect, and the avalanche distributions.
Therefore, investigations on the spatially extended Prisoner's dilemma, which has become a widely used toy model for the emergence of cooperation, have to take into account the stability of possible equilibria depending on chosen payoff matrix, strategy space, and topology. Complex behavior should only be found for subcritical or critical dynamics whereas in the supercritical regime small perturbations will totally mix up the whole system preventing the evolution of local structures.
The results on the stability of the Nash equilibria and their connection to evolutionarily stable states indicate that the concept of equilibrium, originating from classical mechanics and brought into the fields of game theory and evolution \cite{ekeland:2002}, has to be further specified to take into account co-evolution on networks and other spatial structures.

\begin{acknowledgements}
We thank Christoph Hauert, Thomas Lux, and Uwe R\"{o}sler for useful discussions and comments. H.~E.\ gratefully acknowledges support by the Studienstiftung des deutschen Volkes (German National Merit Foundation).
\end{acknowledgements}


\begin{thebibliography}{40}
\expandafter\ifx\csname natexlab\endcsname\relax\def\natexlab#1{#1}\fi
\expandafter\ifx\csname bibnamefont\endcsname\relax
  \def\bibnamefont#1{#1}\fi
\expandafter\ifx\csname bibfnamefont\endcsname\relax
  \def\bibfnamefont#1{#1}\fi
\expandafter\ifx\csname citenamefont\endcsname\relax
  \def\citenamefont#1{#1}\fi
\expandafter\ifx\csname url\endcsname\relax
  \def\url#1{\texttt{#1}}\fi
\expandafter\ifx\csname urlprefix\endcsname\relax\def\urlprefix{URL }\fi
\providecommand{\bibinfo}[2]{#2}
\providecommand{\eprint}[2][]{\url{#2}}

\bibitem[{\citenamefont{Arthur}(1999)}]{arthur:1999}
\bibinfo{author}{\bibfnamefont{W.~B.} \bibnamefont{Arthur}},
  \bibinfo{journal}{Science} \textbf{\bibinfo{volume}{284}},
  \bibinfo{pages}{107} (\bibinfo{year}{1999}).

\bibitem[{\citenamefont{Kirman}(1989)}]{kirman:1989}
\bibinfo{author}{\bibfnamefont{A.}~\bibnamefont{Kirman}}, \bibinfo{journal}{The
  Economic Journal} \textbf{\bibinfo{volume}{99}}, \bibinfo{pages}{126}
  (\bibinfo{year}{1989}).

\bibitem[{\citenamefont{Neumann and
  Morgenstern}(1953)}]{neumann/morgenstern:1953}
\bibinfo{author}{\bibfnamefont{J.~V.} \bibnamefont{Neumann}} \bibnamefont{and}
  \bibinfo{author}{\bibfnamefont{O.}~\bibnamefont{Morgenstern}},
  \emph{\bibinfo{title}{Theory of games and economic behavior}}
  (\bibinfo{publisher}{Princeton University Press},
  \bibinfo{address}{Princeton}, \bibinfo{year}{1953}).

\bibitem[{\citenamefont{Maynard~Smith}(1982)}]{maynard-smith:1982}
\bibinfo{author}{\bibfnamefont{J.}~\bibnamefont{Maynard~Smith}},
  \emph{\bibinfo{title}{Evolution and the theory of games}}
  (\bibinfo{publisher}{Cambridge University Press},
  \bibinfo{address}{Cambridge}, \bibinfo{year}{1982}).

\bibitem[{\citenamefont{Blume}(1995)}]{blume-l:1993}
\bibinfo{author}{\bibfnamefont{L.~E.} \bibnamefont{Blume}},
  \bibinfo{journal}{Games and Economic Behavior} \textbf{\bibinfo{volume}{11}},
  \bibinfo{pages}{111} (\bibinfo{year}{1995}).

\bibitem[{\citenamefont{Nowak et~al.}(1994{\natexlab{a}})\citenamefont{Nowak,
  Latane, and Lewenstein}}]{nowak/lewenstein:1994}
\bibinfo{author}{\bibfnamefont{A.}~\bibnamefont{Nowak}},
  \bibinfo{author}{\bibfnamefont{B.}~\bibnamefont{Latane}}, \bibnamefont{and}
  \bibinfo{author}{\bibfnamefont{M.}~\bibnamefont{Lewenstein}}, in
  \emph{\bibinfo{booktitle}{Social Dilemmas and Cooperation}}, edited by
  \bibinfo{editor}{\bibfnamefont{U.}~\bibnamefont{Schulz}},
  \bibinfo{editor}{\bibfnamefont{W.}~\bibnamefont{Albers}}, \bibnamefont{and}
  \bibinfo{editor}{\bibfnamefont{U.}~\bibnamefont{Mueller}}
  (\bibinfo{publisher}{Springer}, \bibinfo{address}{Berlin},
  \bibinfo{year}{1994}{\natexlab{a}}).

\bibitem[{\citenamefont{Thompson and
  Cunningham}(2002)}]{thompson/cunningham:2002}
\bibinfo{author}{\bibfnamefont{J.~N.} \bibnamefont{Thompson}} \bibnamefont{and}
  \bibinfo{author}{\bibfnamefont{B.~M.} \bibnamefont{Cunningham}},
  \bibinfo{journal}{Nature} \textbf{\bibinfo{volume}{417}},
  \bibinfo{pages}{735} (\bibinfo{year}{2002}).

\bibitem[{\citenamefont{Oborny et~al.}(2000)\citenamefont{Oborny, Kun,
  Cz{\'{a}}r{\'{a}}n, and Bokros}}]{oborny/bokros:2000}
\bibinfo{author}{\bibfnamefont{B.}~\bibnamefont{Oborny}},
  \bibinfo{author}{\bibfnamefont{A.}~\bibnamefont{Kun}},
  \bibinfo{author}{\bibnamefont{Cz{\'{a}}r{\'{a}}n}}, \bibnamefont{and}
  \bibinfo{author}{\bibfnamefont{S.}~\bibnamefont{Bokros}},
  \bibinfo{journal}{Ecology} \textbf{\bibinfo{volume}{81}},
  \bibinfo{pages}{3291} (\bibinfo{year}{2000}).

\bibitem[{\citenamefont{Huberman and Lukose}(1997)}]{huberman/lukose:1997}
\bibinfo{author}{\bibfnamefont{B.~A.} \bibnamefont{Huberman}} \bibnamefont{and}
  \bibinfo{author}{\bibfnamefont{R.~M.} \bibnamefont{Lukose}},
  \bibinfo{journal}{Science} \textbf{\bibinfo{volume}{277}},
  \bibinfo{pages}{535} (\bibinfo{year}{1997}).

\bibitem[{\citenamefont{Miller et~al.}(2002)\citenamefont{Miller, Butts, and
  Rode}}]{miller/rode:2002}
\bibinfo{author}{\bibfnamefont{J.~H.} \bibnamefont{Miller}},
  \bibinfo{author}{\bibfnamefont{C.~T.} \bibnamefont{Butts}}, \bibnamefont{and}
  \bibinfo{author}{\bibfnamefont{D.}~\bibnamefont{Rode}},
  \bibinfo{journal}{Journal of Economic Behavior \& Organization}
  \textbf{\bibinfo{volume}{47}}, \bibinfo{pages}{179} (\bibinfo{year}{2002}).

\bibitem[{\citenamefont{Axelrod}(1984)}]{axelrod:1984}
\bibinfo{author}{\bibfnamefont{R.~K.} \bibnamefont{Axelrod}},
  \emph{\bibinfo{title}{The evolution of cooperation}}
  (\bibinfo{publisher}{Basic Books}, \bibinfo{address}{New York},
  \bibinfo{year}{1984}).

\bibitem[{\citenamefont{Schulz et~al.}(1994)\citenamefont{Schulz, Albers, and
  Mueller}}]{schulz/mueller:1994}
\bibinfo{editor}{\bibfnamefont{U.}~\bibnamefont{Schulz}},
  \bibinfo{editor}{\bibfnamefont{W.}~\bibnamefont{Albers}}, \bibnamefont{and}
  \bibinfo{editor}{\bibfnamefont{U.}~\bibnamefont{Mueller}}, eds.,
  \emph{\bibinfo{title}{Social Dilemmas and Cooperation}}
  (\bibinfo{publisher}{Springer}, \bibinfo{address}{Berlin},
  \bibinfo{year}{1994}).

\bibitem[{\citenamefont{Axelrod and Hamilton}(1981)}]{axelrod/hamilton:1981}
\bibinfo{author}{\bibfnamefont{R.}~\bibnamefont{Axelrod}} \bibnamefont{and}
  \bibinfo{author}{\bibfnamefont{W.~D.} \bibnamefont{Hamilton}},
  \bibinfo{journal}{Science} \textbf{\bibinfo{volume}{211}},
  \bibinfo{pages}{1390} (\bibinfo{year}{1981}).

\bibitem[{\citenamefont{Nowak and May}(1992)}]{nowak/may:1992}
\bibinfo{author}{\bibfnamefont{M.~A.} \bibnamefont{Nowak}} \bibnamefont{and}
  \bibinfo{author}{\bibfnamefont{R.~M.} \bibnamefont{May}},
  \bibinfo{journal}{Nature} \textbf{\bibinfo{volume}{359}},
  \bibinfo{pages}{826} (\bibinfo{year}{1992}), \bibinfo{note}{52}.

\bibitem[{\citenamefont{Nowak and May}(1993)}]{nowak/may:1993}
\bibinfo{author}{\bibfnamefont{M.~A.} \bibnamefont{Nowak}} \bibnamefont{and}
  \bibinfo{author}{\bibfnamefont{R.~M.} \bibnamefont{May}},
  \bibinfo{journal}{Int. J. Bifurcation Chaos} \textbf{\bibinfo{volume}{3}},
  \bibinfo{pages}{35} (\bibinfo{year}{1993}).

\bibitem[{\citenamefont{Huberman and Glance}(1993)}]{huberman/glance:1993}
\bibinfo{author}{\bibfnamefont{B.~A.} \bibnamefont{Huberman}} \bibnamefont{and}
  \bibinfo{author}{\bibfnamefont{N.~S.} \bibnamefont{Glance}},
  \bibinfo{journal}{Proc. Natl. Acad. Sci. USA} \textbf{\bibinfo{volume}{90}},
  \bibinfo{pages}{7716} (\bibinfo{year}{1993}).

\bibitem[{\citenamefont{Nowak et~al.}(1994{\natexlab{b}})\citenamefont{Nowak,
  Bonhoeffer, and May}}]{nowak/may:1994}
\bibinfo{author}{\bibfnamefont{M.~A.} \bibnamefont{Nowak}},
  \bibinfo{author}{\bibfnamefont{S.}~\bibnamefont{Bonhoeffer}},
  \bibnamefont{and} \bibinfo{author}{\bibfnamefont{R.~M.} \bibnamefont{May}},
  \bibinfo{journal}{Proc. Natl. Acad. Sci. USA} \textbf{\bibinfo{volume}{91}},
  \bibinfo{pages}{4877} (\bibinfo{year}{1994}{\natexlab{b}}).

\bibitem[{\citenamefont{Nowak et~al.}(1994{\natexlab{c}})\citenamefont{Nowak,
  Bonhoeffer, and May}}]{nowak/may:1994b}
\bibinfo{author}{\bibfnamefont{M.~A.} \bibnamefont{Nowak}},
  \bibinfo{author}{\bibfnamefont{S.}~\bibnamefont{Bonhoeffer}},
  \bibnamefont{and} \bibinfo{author}{\bibfnamefont{R.~M.} \bibnamefont{May}},
  \bibinfo{journal}{Int. J. Bifurcation Chaos} \textbf{\bibinfo{volume}{4}},
  \bibinfo{pages}{33} (\bibinfo{year}{1994}{\natexlab{c}}).

\bibitem[{\citenamefont{Herz}(1994)}]{herz:1994}
\bibinfo{author}{\bibfnamefont{A.~V.~M.} \bibnamefont{Herz}},
  \bibinfo{journal}{J. Theor. Biol.} \textbf{\bibinfo{volume}{169}},
  \bibinfo{pages}{65} (\bibinfo{year}{1994}).

\bibitem[{\citenamefont{Szab{\'{o}} et~al.}(2000)\citenamefont{Szab{\'{o}},
  Antal, Szab{\'{o}}, and Droz}}]{szabo/droz:2000}
\bibinfo{author}{\bibfnamefont{G.}~\bibnamefont{Szab{\'{o}}}},
  \bibinfo{author}{\bibfnamefont{T.}~\bibnamefont{Antal}},
  \bibinfo{author}{\bibfnamefont{P.}~\bibnamefont{Szab{\'{o}}}},
  \bibnamefont{and} \bibinfo{author}{\bibfnamefont{M.}~\bibnamefont{Droz}},
  \bibinfo{journal}{Phys. Rev. E.} \textbf{\bibinfo{volume}{62}},
  \bibinfo{pages}{1095} (\bibinfo{year}{2000}).

\bibitem[{\citenamefont{Szab{\'{o}} and T{\H{o}}ke}(1998)}]{szabo/toke:1998}
\bibinfo{author}{\bibfnamefont{G.}~\bibnamefont{Szab{\'{o}}}} \bibnamefont{and}
  \bibinfo{author}{\bibfnamefont{C.}~\bibnamefont{T{\H{o}}ke}},
  \bibinfo{journal}{Phys.\ Rev.\ E} \textbf{\bibinfo{volume}{58}},
  \bibinfo{pages}{69} (\bibinfo{year}{1998}).

\bibitem[{\citenamefont{Chiappin and
  de~Oliveira}(1999)}]{chiappin/oliveira:1999}
\bibinfo{author}{\bibfnamefont{J.~R.~N.} \bibnamefont{Chiappin}}
  \bibnamefont{and} \bibinfo{author}{\bibfnamefont{M.~J.}
  \bibnamefont{de~Oliveira}}, \bibinfo{journal}{Phys. Rev. E}
  \textbf{\bibinfo{volume}{59}}, \bibinfo{pages}{6419} (\bibinfo{year}{1999}).

\bibitem[{\citenamefont{Vainstein and Arenzon}(2001)}]{vainstein/arenzon:2001}
\bibinfo{author}{\bibfnamefont{M.~H.} \bibnamefont{Vainstein}}
  \bibnamefont{and} \bibinfo{author}{\bibfnamefont{J.~J.}
  \bibnamefont{Arenzon}}, \bibinfo{journal}{Phys.\ Rev.\ E}
  \textbf{\bibinfo{volume}{64}}, \bibinfo{pages}{051905} (\bibinfo{year}{2001}).

\bibitem[{\citenamefont{Tomochi and Kono}(2002)}]{tomochi/kono:2002}
\bibinfo{author}{\bibfnamefont{M.}~\bibnamefont{Tomochi}} \bibnamefont{and}
  \bibinfo{author}{\bibfnamefont{M.}~\bibnamefont{Kono}},
  \bibinfo{journal}{Phys.\ Rev.\ E} \textbf{\bibinfo{volume}{65}},
  \bibinfo{pages}{026112} (\bibinfo{year}{2002}).

\bibitem[{\citenamefont{Lim et~al.}(2002)\citenamefont{Lim, Chen, and
  Jayaprakash}}]{lim/jayaprakash:2002}
\bibinfo{author}{\bibfnamefont{Y.~F.} \bibnamefont{Lim}},
  \bibinfo{author}{\bibfnamefont{K.}~\bibnamefont{Chen}}, \bibnamefont{and}
  \bibinfo{author}{\bibfnamefont{C.}~\bibnamefont{Jayaprakash}}, \bibinfo{journal}{Phys.\ Rev.\
  E} \textbf{\bibinfo{volume}{65}}, \bibinfo{pages}{026134}
  (\bibinfo{year}{2002}).

\bibitem[{\citenamefont{Lindgren and Nordahl}(1994)}]{lindgren/nordahl:1994}
\bibinfo{author}{\bibfnamefont{K.}~\bibnamefont{Lindgren}} \bibnamefont{and}
  \bibinfo{author}{\bibfnamefont{M.~G.} \bibnamefont{Nordahl}},
  \bibinfo{journal}{Physica D} \textbf{\bibinfo{volume}{75}},
  \bibinfo{pages}{292} (\bibinfo{year}{1994}).

\bibitem[{\citenamefont{Hauert and Schuster}(1997)}]{hauert/schuster:1997}
\bibinfo{author}{\bibfnamefont{C.}~\bibnamefont{Hauert}} \bibnamefont{and}
  \bibinfo{author}{\bibfnamefont{H.~G.} \bibnamefont{Schuster}},
  \bibinfo{journal}{Proc. R. Soc. Lond. B} \textbf{\bibinfo{volume}{264}},
  \bibinfo{pages}{513} (\bibinfo{year}{1997}).

\bibitem[{\citenamefont{Hauert and Schuster}(1998)}]{hauert/schuster:1998}
\bibinfo{author}{\bibfnamefont{C.}~\bibnamefont{Hauert}} \bibnamefont{and}
  \bibinfo{author}{\bibfnamefont{H.~G.} \bibnamefont{Schuster}},
  \bibinfo{journal}{J. Theor. Biol.} \textbf{\bibinfo{volume}{192}},
  \bibinfo{pages}{155} (\bibinfo{year}{1998}).

\bibitem[{\citenamefont{Hauert}(1999)}]{hauert:1999}
\bibinfo{author}{\bibfnamefont{C.}~\bibnamefont{Hauert}},
  \emph{\bibinfo{title}{Evolution of cooperation -- The Prisoner's dilemma and
  its applications as an example.}} (\bibinfo{publisher}{Shaker},
  \bibinfo{address}{Aachen}, \bibinfo{year}{1999}).

\bibitem[{\citenamefont{Abramson and Kuperman}(2001)}]{abramson/kuperman:2001}
\bibinfo{author}{\bibfnamefont{G.}~\bibnamefont{Abramson}} \bibnamefont{and}
  \bibinfo{author}{\bibfnamefont{M.}~\bibnamefont{Kuperman}},
  \bibinfo{journal}{Phys.\ Rev.\ E} \textbf{\bibinfo{volume}{63}},
  \bibinfo{pages}{030901(R)} (\bibinfo{year}{2001}).

\bibitem[{\citenamefont{Zimmermann et~al.}(2001)\citenamefont{Zimmermann,
  Egu{\'{i}}luz, and Miguel}}]{zimmermann/miguel:2001}
\bibinfo{author}{\bibfnamefont{M.~G.} \bibnamefont{Zimmermann}},
  \bibinfo{author}{\bibfnamefont{V.~M.} \bibnamefont{Egu{\'{i}}luz}},
  \bibnamefont{and} \bibinfo{author}{\bibfnamefont{M.~S.}
  \bibnamefont{Miguel}}, in \emph{\bibinfo{booktitle}{Economics with
  heterogeneous interacting agents}}, edited by
  \bibinfo{editor}{\bibfnamefont{A.}~\bibnamefont{Kirman}} \bibnamefont{and}
  \bibinfo{editor}{\bibfnamefont{J.-B.} \bibnamefont{Zimmermann}}
  (\bibinfo{publisher}{Springer}, \bibinfo{address}{Berlin},
  \bibinfo{year}{2001}).

\bibitem[{\citenamefont{Fudenberg and Tirole}(1998)}]{fudenberg/tirole:1998}
\bibinfo{author}{\bibfnamefont{D.}~\bibnamefont{Fudenberg}} \bibnamefont{and}
  \bibinfo{author}{\bibfnamefont{J.}~\bibnamefont{Tirole}},
  \emph{\bibinfo{title}{Game theory}} (\bibinfo{publisher}{MIT Press},
  \bibinfo{address}{Cambridge, London}, \bibinfo{year}{1998}).

\bibitem[{\citenamefont{Nash}(1950)}]{nash:1950}
\bibinfo{author}{\bibfnamefont{J.~F.} \bibnamefont{Nash}},
  \bibinfo{journal}{Proc. Natl. Acad. Sci. USA} \textbf{\bibinfo{volume}{36}},
  \bibinfo{pages}{48} (\bibinfo{year}{1950}).

\bibitem[{\citenamefont{Weibull}(1996)}]{weibull:1996}
\bibinfo{author}{\bibfnamefont{J.}~\bibnamefont{Weibull}},
  \emph{\bibinfo{title}{Evolutionary Game Theory}} (\bibinfo{publisher}{MIT
  Press}, \bibinfo{address}{Cambridge, London}, \bibinfo{year}{1996}).

\bibitem[{\citenamefont{Hofbauer and Sigmund}(1998)}]{hofbauer/sigmund:1998}
\bibinfo{author}{\bibfnamefont{J.}~\bibnamefont{Hofbauer}} \bibnamefont{and}
  \bibinfo{author}{\bibfnamefont{K.}~\bibnamefont{Sigmund}},
  \emph{\bibinfo{title}{Evolutionary Games and Population Dynamics}}
  (\bibinfo{publisher}{Cambridge University Press},
  \bibinfo{address}{Cambridge}, \bibinfo{year}{1998}).

\bibitem[{\citenamefont{Harris}(1963)}]{harris:1963}
\bibinfo{author}{\bibfnamefont{T.~E.} \bibnamefont{Harris}},
  \emph{\bibinfo{title}{The theory of branching processes}}
  (\bibinfo{publisher}{Springer}, \bibinfo{address}{Berlin},
   \bibinfo{year}{1963}).

\bibitem[{\citenamefont{Jagers}(1975)}]{jagers:1975}
\bibinfo{author}{\bibfnamefont{P.}~\bibnamefont{Jagers}},
  \emph{\bibinfo{title}{Branching processes with biological applications}}
  (\bibinfo{publisher}{Wiley}, \bibinfo{address}{London}, \bibinfo{year}{1975}).

\bibitem[{\citenamefont{Bak}(1996)}]{bak:1996}
\bibinfo{author}{\bibfnamefont{P.}~\bibnamefont{Bak}},
  \emph{\bibinfo{title}{How nature works: the science of self-organized
  criticality}} (\bibinfo{publisher}{Springer}, \bibinfo{address}{New York},
  \bibinfo{year}{1996}).

\bibitem[{\citenamefont{Jensen}(1998)}]{jensen-x:1998}
\bibinfo{author}{\bibfnamefont{H.~J.} \bibnamefont{Jensen}},
  \emph{\bibinfo{title}{Self-organized criticality: Emergent behavior in
  physical and biological systems}} (\bibinfo{publisher}{Cambridge University
  Press}, \bibinfo{address}{Cambridge}, \bibinfo{year}{1998}).

\bibitem[{\citenamefont{Ekeland}(2002)}]{ekeland:2002}
\bibinfo{author}{\bibfnamefont{I.}~\bibnamefont{Ekeland}},
  \bibinfo{journal}{Nature} \textbf{\bibinfo{volume}{417}},
  \bibinfo{pages}{385} (\bibinfo{year}{2002}).

\end{thebibliography}
\end{document}